\documentclass[11pt]{article}
\usepackage{amsmath,amsthm,amscd,amssymb}
\usepackage{latexsym}
\usepackage{graphicx,epsfig}
\makeindex
\newcommand{\be}{\begin{equation}}
\newcommand{\ee}{\end{equation}}
\newcommand{\bea}{\begin{eqnarray}}
\newcommand{\eea}{\end{eqnarray}}

\def\asec{$''$ cy$^{-1}$}

\def\bb{\bibitem}
\def\rfr#1{eq.(\ref{#1})}

\def\eqi{\begin{equation}}
\def\eqf{\end{equation}}
\begin{document}  
\begin{titlepage} 
\begin{flushright}
\today\\
\end{flushright}
\vspace{.5cm}
\begin{center}
{\LARGE Can Solar System observations tell us something about the cosmological constant?\\}
\vspace{0.5cm}
\quad\\
{Lorenzo Iorio\\
\vspace{0.5cm}
\quad\\
Viale Unit\`a di Italia 68, 70125,
Bari, Italy\\ \vspace{0.5cm}
}
\vspace*{0.5cm}

{\bf Abstract\\}
\end{center}
In this note we show that the latest determinations of the residual perihelion advances  $\delta\dot\varpi$ of the inner planets of the Solar System, obtained by accounting for almost all known Newtonian and post-Newtonian orbital effects, yield only very broad constraints on the cosmological constant $\Lambda$. Indeed, from, e.g.,  $\delta\dot\varpi=-0.0036\pm 0.0050$ arcseconds per century for Mercury one gets 
$-2\times 10^{-34}\ {\rm km}^{-2} < \Lambda < 4\times 10^{-35}\ {\rm km}^{-2}.$ The currently accepted value for $\Lambda$, obtained from many independent cosmological and large-scale measurements, amounts to almost $10^{-46}$ km$^{-2}$.
{\noindent \small  } \end{titlepage}
\newpage
\setcounter{page}{1}
\vspace{0.2cm}
\baselineskip 14pt

\setcounter{footnote}{0}
\setlength{\baselineskip}{1.5\baselineskip}
\renewcommand{\theequation}{\mbox{$\arabic{equation}$}}
\noindent

\section{Introduction}
 The possibility of constraining the cosmological constant $\Lambda$ by means of  observations of Mercury's perihelion advance was investigated by Cardona and Tejeiro \cite{Cardo}  in the framework of the usual four-dimensional general relativity. They used results accurate to 0.1 arcseconds per century (\asec\ in the following) to derive a bound
 $|\Lambda|<10^{-45}$ km$^{-2}$. The current accepted value for $\Lambda$, obtained from many independent large-scale and cosmological measurements, amounts to almost $10^{-46}$ km$^{-2}$. The authors of \cite{Cardo} claim that an improvement of one-two orders of magnitude in the precision of the Mercury's orbit determination could allow to yield bounds to $\Lambda$ competitive with those of cosmological origin. Other researchers have been influenced by such conclusions. E.g., Dumin \cite{Dumin} starts from them claiming, among other things, that, up to now, better measurements of the Mercury's perihelion shift are not yet available.

In this note we will show that the improvements in planetary orbit data reduction hoped by Cardona and Tejeiro, in fact, recently occurred but they are quite far from allowing to satisfactorily bound the cosmological constant.

\section{The latest determinations of the planetary perihelia and the precession induced by the cosmological constant}
Pitjeva recently processed more than 317 000 observations (1913-2003) of various types including radiometric measurements of planets and spacecraft, astrometric CCD observations of the outer planets and their satellites, and meridian
and photographic observations to construct the EPM2004 ephemerides \cite{Pit05a}. In the adopted dynamical force models there are all the Newtonian N-body features of motion, including also the impact of the largest 301 asteroids and of the asteroid ring that lies in the ecliptic plane \cite{Kraetal02}, the Newtonian effect of the solar oblateness $J_2$ \cite{J2} and the post-Newtonian general relativistic gravitoelectric forces \cite{Einstein 1915}, expressed in terms of the PPN parameters \cite{Will93} $\beta$ and $\gamma$. Neither the general relativistic gravitomagnetic forces \cite{Lense and Thirring 1918} nor any post-Einsteinian effect are included, so that the obtained residuals account entirely for them, if they exist in nature.
In a particular solution in which the PPN parameters and the solar quadrupole mass moment were held fixed to their reference values $\beta=\gamma=1,\ J_2=2\times 10^{-7}$,  Pitjeva \cite{Pit05b} included  also the secular rates of the longitudes of perihelia in the set of the almost 200  simultaneously fitted parameters. Their determined values are reported in Table 3 of \cite{Pit05b} part of which is reproduced here in Table \ref{Pit}. It is important to note that the quoted uncertainties are not the mere formal, statistical errors but are realistic in the sense that they were obtained from comparison of many different
solutions with different sets of parameters and observations (Pitjeva, private communication 2005a). The correlations among such determined planetary perihelia rates are very low with a maximum of about $20\%$ between Mercury and the Earth (Pitjeva, private communication 2005b). 
\begin{table}
\caption{Determined extra-secular precessions of the longitudes of perihelia of the inner planets of the Solar System, from Table 3 of \cite{Pit05b}. The units used are \asec. Note that the eccentricity of Venus amounts to 0.0066 only, so that its perihelion is not a good observable. The quoted errors are not the formal, statistical ones and the correlations among the determined perihelia rates are very low, with a maximum of about $20\%$ between Mercury and the Earth.}\label{Pit}
\begin{tabular}{@{\hspace{0pt}}cccc}
\hline\noalign{\smallskip} Mercury & Venus & Earth & Mars\\
\noalign{\smallskip}\hline\noalign{\smallskip} 
$-0.0036\pm 0.0050$ & $0.53\pm 0.30$ & $-0.0002\pm 0.0004$ & $0.0001\pm 0.0005$\\
 \noalign{\smallskip}\hline
\end{tabular}
\end{table} 

A useful explicit expression of the perihelion precession induced by $\Lambda$ was derived in \cite{Mash}.
It is
\eqi\dot\varpi_{\Lambda}=\frac{1}{2}\frac{\Lambda c^2}{n}\sqrt{1-e^2}\label{pre},\eqf
where $n=\sqrt{GM/a^3}$ is the Keplerian mean motion ($G$ is the gravitational constant, $M$ is the mass of the central body and $a$ is the semimajor axis of the orbit) and $e$ is the orbital eccentricity.
It is important to note that, contrary to (6) in \cite{Cardo}, \rfr{pre} was obtained by using a radial isotropic coordinate, which is commonly used in the planetary data reductions.

The impossibility of using the Solar System data becomes apparent if we calculate the magnitude of the precessions of  
\rfr{pre} for the inner planets: indeed, they amount to about  $10^{-14}-10^{-15}$ \asec. If we straightforwardly use the results  of 
Table \ref{Pit} and \rfr{pre} for the inner planets to constrain the cosmological constant we find
the results of Table \ref{Lamc}.
\begin{table}
\caption{Bounds on $\Lambda$, in km$^{-2}$, from the data of Table \ref{Pit}.}\label{Lamc}
\begin{tabular}{@{\hspace{0pt}}ccccc}
\hline\noalign{\smallskip} & Mercury & Venus & Earth & Mars\\
\noalign{\smallskip}\hline\noalign{\smallskip} 
$\Lambda_{\rm min}$ & $-2\times 10^{-34}$ & $2\times 10^{-33}$ & $-4\times 10^{-36}$ & $-2\times 10^{-36}$ \\
$\Lambda_{\rm max}$ & $4\times 10^{-35}$ & $9\times 10^{-33}$ & $1\times 10^{-36}$ & $1\times 10^{-36}$ \\
 \noalign{\smallskip}\hline
\end{tabular}
\end{table} 

As shown in \cite{buu}, the effect
of $\Lambda$ starts to become significant if Megaparsec distances are
considered. 

\section{Conclusions}
In this note we have discussed the possibility of constraining the cosmological constant $\Lambda$, in a general relativistic framework, with Solar System observations in view of the latest results in planetary orbit determinations.
Contrary to what claimed by some authors, it turns out that it is not possible to get useful bounds on $\Lambda$ from such local scale tests.

\section*{Acknowledgements}
I thank E.V. Pitjeva and C.G B$\ddot{\rm o}$hmer for useful correspondence.

\end{document}